# Neutral and Stable Equilibria of Genetic Systems and The Hardy-Weinberg Principle: Limitations of the Chi-Square Test and Advantages of Auto-Correlation Functions of Allele Frequencies

Francisco Bosco[1,2], Diogo Castro[3] and Marcelo R. S. Briones[1,2]

[1]Departamento de Microbiologia, Imunologia e Parasitologia, Universidade Federal de São Paulo, Rua Botucatu 862, ECB 3º andar, CEP 04023-062, São Paulo, SP, Brazil, [2]Laboratório de Genômica Evolutiva e Biocomplexidade, Universidade Federal de São Paulo, Rua Pedro de Toledo 669, 4º andar, CEP 04039-032, São Paulo, SP, Brazil, [3]Departamento de Medicina, Disciplina de Infectologia, Universidade Federal de São Paulo, Rua Botucatu 740, CEP 04023-900, São Paulo, SP, Brazil.

**Abstract**

Since the foundations of Population Genetics the notion of genetic equilibrium (in close analogy to Classical Mechanics) has been associated to the Hardy-Weinberg (HW) Principle and the identification of equilibrium is currently assumed by stating that the HW axioms are valid if appropriate values of $X^2$ ($p<0.05$) are observed in experiments. Here we show by numerical experiments with the genetic system of one locus/two alleles that considering large ensembles of populations the $X^2$-test is not decisive and may lead to false negatives in random mating populations and false positives in nonrandom mating populations. As a result we confirm the logical statement that statistical tests can not be used to deduce if the genetic population is under the HW conditions. Furthermore, we show that under the HW conditions populations of any finite size evolve in time according to what can be identified as neutral dynamics to which the very notion of equilibrium is unattainable for any practical purpose. Therefore, under the HW conditions equilibrium properties are not observable. We also show that by relaxing the condition of random mating the dynamics acquires all the characteristics of asymptotic stable equilibrium. As a consequence our results show that the question of equilibrium in genetic systems should be approached in close analogy to non-equilibrium statistical physics and its observability should be focused on dynamical quantities like the typical decay properties of the allelic auto correlation function in time. In this perspective one should abandon the classical notion of genetic equilibrium and its relation to the HW proportions and open investigations in the direction of searching for unifying general principles of population genetic transformations capable to take in consideration these systems in their full complexity.

**Corresponding author:** Marcelo R.S. Briones
Departamento de Microbiologia, Imunologia e Parasitologia,
Universidade Federal de São Paulo,
Rua Botucatu 862, ECB 3 andar, CEP 04023-062, São Paulo, SP, Brazil
TEL: 55 11 5083-2980, FAX: 55 11 5572-4711,
EMAIL: marcelo.briones@unifesp.br

**Introduction**

In his letter to the Editor of Science in 1908, G. Hardy (Hardy, 1908) showed that under specific conditions the simple two allele system (A,a) has the property that the allele frequencies ( $p,q$ ) determine the genotype frequencies (AA, aa, Aa) obeying proportions given by the simple well known relation $p^2+q^2+2\mathrm{pq}=1$ . Independently, the German physician W. Weinberg (Weinberg, 1908) arrived to similar results. In the 1930's the synthetic theory proposed that under the Hardy-Weinberg (HW) conditions genetic systems attain an "equilibrium" state characterized by the genotype frequency proportions obtained directly from the allele frequencies as stated in the HW relation above. This definition of equilibrium in genetic systems became part of the well known HW Principle (Hartl and Clarke, 2007) which formally states that: "If a genetic population is such that (1) organisms are diploid, (2) reproduction is sexual, (3) generations do not overlap, (4) matings are random, (5) the size of the population is significantly large, (6) allele frequencies are equal in the sexes and (7) there is no migration, mutation or selection, then the genotype frequencies in the population are given by weighted products of the allele frequencies. In the case of the one locus-two allele system the allele frequencies ( $(A,a)=(p,q)$ ) give directly the genotype frequencies ( $(AA,aa,Aa)=(p^2,q^2,2\mathrm{pq})$ )." Under the above conditions it is easy to demonstrate the following corollary: "For a population satisfying the HW conditions the allele frequencies are constant in time". The notion of HW equilibrium adopted in the synthetic theory derives from the above corollary.

Attempting to establish a conceptual "bridge" with the definition of mechanical equilibrium in Newtonian mechanics, the synthetic theory adopts the idea that the existence of a time invariant observable leads to the hypothesis that the genetic population would be in equilibrium with no net "external forces" acting on the system (Stephens, 2004; Hartl and Clarke, 2007). From this simple reasoning it became broadly accepted that these external forces should be represented by selection and therefore the theory derives the well known definition of evolution based on the variation of the allele frequencies. This vision of biological evolution is strongly supported by the well defined concept of mechanical equilibrium of (classical) physical systems: if the sum of all external forces is null, the physical system is said to be isolated and its (macroscopic) mechanical state (well characterized by appropriate mechanical variables of its constituents) is unaltered. In this situation the system is said to be in mechanical equilibrium. This notion is a result of the combined use of the first and second Newton's principles.

In the case of genetic systems, assuming that selection "forces" are not acting on the population, the system would be free of external forces and would be in a "genetic state" for which the allele frequencies are constant in time. The idea of genetic equilibrium state under zero "external forces" would follow immediately. Although this theoretical construct may be appealing it is important to note that it is no more than an analogy with serious difficulties to be formally established as it is done in classical mechanics: in classical mechanics the notion of a force acting on a system (as a result of fundamental physical interaction) has the formal status of a (operational) definition. In fact, from the formal point of view, the first two Newtonian Principles are definitions. Only the third principle, the Action-Reaction Principle, has a more fundamental role since it is related to the (universal) law of conservation of momentum and intrinsic symmetries of the physical system. The formal results of classical mechanics can not be derived solely by the first two principles. The third law is in the very foundations of the theory since it addresses the physical phenomenon itself (the interaction of two physical systems obeys basic principles).

In genetic systems viewed in analogy with classical mechanical systems, natural selection appears as the definition of a cause external to the system and capable of changing the system's state, in a clear analogy to the second principle of Newtonian mechanics. Therefore, if the analogy is fully considered, natural selection should be placed as a formal definition. In this line of thought, the biological law analogous to the third law of classical mechanics is still to be found.

The above arguments should be sufficient to understand that the notion of genetic equilibrium in the framework of the synthetic theory is (at most) related to the concept of statistical equilibrium. Its relevance as a natural principle should be supported by experiments. Also, as a matter o fact, in his original study, G. Hardy (Hardy, 1908) carefully chose "stability" and not equilibrium to describe the statistical invariance of allele frequencies. This is more than a semantics difference because a system can be stable without being at equilibrium, as for example is the case of metastable states in non-equilibrium thermodynamics and therefore stability and equilibrium do not necessarily refer to the same physical properties of dynamic systems (Kivelson, 1999).

As it appears in textbooks on population genetics the canonical $X^2$ statistical test is used to compare observed and estimated proportions of alleles (Hartl and Clarke, 2007). The basic idea is that if the measured number of alleles in the population is statistically close enough ( $X^2 < 3.8414...$ and correspondingly $p < 0.05$ ) to the theoretically expected (given by the HW proportions) then the population is said to be in HW equilibrium; in other words the $X^2$ test is used as an indicator of deviation from randomness. Nevertheless, as stated in the HW theorem the set of properties the population has to satisfy constitute necessary but not sufficient conditions for the invariance of the allele frequencies. Therefore, in rigorous terms by merely satisfying the statistical condition $X^2 < 3.8414...$ one can not guarantee random mating or any of the other conditions (or premises) of the HW principle. To prove this statement it suffices to find a counterexample, namely: in a genetic system for which at least one of the conditions of the HW theorem is violated it is possible to satisfy the condition $X^2 < 3.8414...$ . Analytical counterexamples and a possible generalization of the HW principle have already been studied by Li (Li, 1988) and Stark (Stark, 1976; Stark, 1980; Stark, 2005; Stark, 2006a; Stark, 2006b).

To exploit more deeply the possibilities of the system through counterexamples and other statistical properties relevant for the basic concept of genetic equilibrium, we performed numerical experiments with the simplest genetic system of 2 alleles and 3 genotypes (AA,Aa,aa). The numerical simulations show that the stable ensemble distribution of $X^2$ leads to inconclusiveness about random/nonrandom mating for populations of any finite size. The constancy of the allele frequencies can only be observed for rigorously infinite populations under strict HW conditions. As a consequence we present strong arguments and evidence supporting the conclusion that for any finite genetic population under the HW conditions the time evolution of the allele frequencies is dynamically neutral with a corresponding equilibrium state attainable only in the infinite time horizon, and therefore not well characterized by isolated observations. On the other hand in the case of nonrandom mating the allele frequencies obeys a stable dynamics with a well characterized stable equilibrium state.

**Methods**

Numerical experiments were performed with a population of $N$ individuals composed by the three subpopulations with $N_{AA}, N_{Aa}, N_{aa}$ individuals such that $N=N_{AA}+N_{Aa}+N_{aa}$. At each time step $N_c$ reproducing couples were chosen from the population according to a prescribed probability distribution assigned to each individual (**Figure 1**). Each generation step was completed when the couples reproduced; the generation step was considered as the discrete time unit of the model.

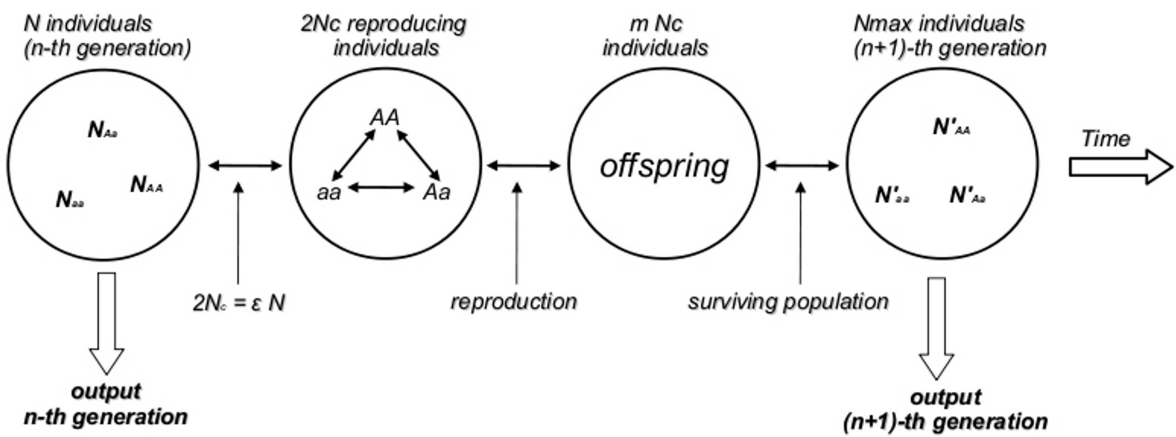

**Figure 1.** Diagram of the numerical experiment that generates the time evolution of the genetic population. This figure illustrates one reproductive cycle. From the $n-th$ generation the algorithm randomly selects $N_c$ couples ($2N_c$ reproducing individuals) corresponding to a fraction ε of the total population $N$. Couples are selected according to specific rules (random or not) and reproduction takes place with a mean number $m$ of descendants per couple. When the population grows at geometric rate ($m>2$) a prescribed number $N_{max}$ of individuals composes the surviving population defined as the $(n+1)-th$ generation.

The case of random mating corresponds to equiprobable individuals; the probabilities to choose any individual are given by the instantaneous finite fractions $f_{ij}=N_{ij}/N$ for each of the genotypes; in the case $N\rightarrow\infty$ and $N_{ij}\rightarrow\infty$ the fractions $f_{ij}$ converge to stable probabilities. For $N\rightarrow\infty$ under the remaining HW conditions the frequencies $f_{ij}$ are time invariant and relates to the HW proportions. For each chosen couple the model specified the number of descendants that could be deterministic and equal for every couple or could be probabilistically chosen from the values 1, 2, 3 or 4. In this case the population had a mean and stable reproductive capacity. Clearly, if the mean reproductive capacity was $m=2$ then the size of the population fluctuated around a constant mean value; if $m>2$ the population size grows geometrically – here (after reproduction) a fixed number of individuals are randomly chosen as viable ones and will reproduce in the next generation; if $m<2$ the population trivially goes to extinction in finite time.

The numerical simulation started by prescribing an initial population with $N$ individuals with the genotype profile $(N_{AA}, N_{Aa}, N_{aa})$ at $(t=0)$; the parameters are the maximum population size $N_{max}$ composed by randomly chosen individuals in case of geometric growth, a set of four stationary probabilities assigned to the four possible values of the reproductive capacity of each couple (determining the mean reproductive capacity $m$), a probability matrix used to prescribe mating chances among the individuals (random mating being a special case) and a second probability matrix that controlled the emergence of mutations. It is important to note that during the process of mating the condition of random mating is strongly dependent on the number of (still) uncoupled individuals; in order to assure random mating for all individuals reproducing in the next generation a fraction of the (already large) population has to be disregarded. In other words, random mating is assured as much as one imposes large values of the reproductive capacity together with large values of death rate. A parameter specifying the reproducing fraction of the population was used to distinguish full and partial mating populations.

As outputs we measured the time evolution of the genotype frequencies, allele frequencies and the value of $X^2$ at each time step. In the analysis we restricted to the cases of finite (although large enough) population effects and non-random mating to analyze the implications for the notion or concept of equilibrium in genetic systems.

The source program was written in C++ on a 64-bits Linux platform. The executable file is available upon request.

**Results**

The $X^2$-test is used in experimental trials with genetic systems as a tool with the specific goal to identify if the genetic system is in a state of equilibrium characterized (and defined) by the conditions stated in the HW principle. This strategy, normally adopted in many studies (Salanti *et al.*, 2005; Rodriguez *et al.*, 2009), would be logically correct if the HW theorem would state necessary and sufficient conditions. But this is not the case. In fact, the statement that is easily proven is: "If the axioms apply then the HW proportions are easily obtained and the time invariance of the allele frequencies follows". The converse can not be proven and therefore the HW axioms imply very small values of $X^2$ (the necessary condition) but very small values of $X^2$ are not enough to guarantee the genetic system is subjected to the HW conditions (the sufficient condition). Therefore the conclusion that the genetic system is in a state of (HW) equilibrium because the measured $X^2 < 3.8414...$ (or any other value) lacks logical foundation.

The $X^2$ value is very sensitive to fluctuations. In fact, its usefulness and limitations for the study of genetic systems are currently under investigation (Rohlfs and Weir, 2008). In numerical experiments with populations up to $10^6 - 10^7$ individuals and obeying the HW conditions it is not difficult to observe values of $X^2 > 5$ or more. It is also true that values below the accepted bound of 3.8414 dominate the time history of the system, but these numerical evidences indicate that finding values of $X^2<3.8414...$ depends on the right moment the measurement takes place; the population may be under the HW conditions but the measured value of $X^2$ is above the accepted threshold (a case of false negative). Therefore the statistical test can not be considered decisive to make a conclusive statement about the state of the genetic system.

The other extreme situation emerges when the population is subject to (for instance) non-random mating and therefore under conditions where the HW principle is not valid. For a large number of different mating probabilities the time history of the system presents a significant number of time periods when $X^2<3.8414$ and therefore actual measurements during these time intervals would produce misleading conclusions (false positives). In logical/mathematical terms these are consequences of the fact that the $X^2$ statistical test can not be used to prove or disprove axiomatic statements.

In order to study the fluctuations of $X^2$ we avoid the problem of sampling by calculating the distribution of possible values of $X^2$ over a statistical ensemble $Z_L(N)$ made up of $L$ copies of identical populations of size $N$. For each population of the ensemble we evolve the system till a fixed time point where the instantaneous value of $X^2$ is measured. As a result we obtain the stable $Z_L(N)$ - ensemble distribution of possible values of $X^2$ for populations of size $N$ at fixed time. In the numerical experiments we consider simulations where $N=3.0 \times 10^6$ and $L=10^5$. Initially all populations of the ensemble are identical with the same genotype distribution $N_{aa}=N_{AA}=N_{Aa}=1.0 \times 10^6$. At each time step the number of reproducing individuals is fixed at $N_{rep}=1.8 \times 10^6$ in such a way that imbalance of the mating probabilities is minimized at each time step. Geometric reproduction rate is imposed to guarantee the same value of $N_{rep}$ at each time step. The measurement of $X^2$ is made for each population of the ensemble at the generation 30. Here we present the results of two typical simulations. **Figure 2** shows the ensemble distributions of $X^2$ for populations under random mating condition and one example of nonrandom mating condition where the mating probabilities are fixed as $P(AA)=0.23, P(aa)=0.23, P(Aa)=0.54$ in order to choose reproducing couples. As it can be clearly observed the two distributions are almost indistinguishable.

Both distributions have a well fitted exponential tail for $X^2>0.4$. With the help of the ensemble distribution it is possible to estimate the probability of false negatives in the case of random mating as $P(X^2>3.8414)=0.146$. We note that this value is very robust in the sense that it is (almost) constant in the range $N=10^4-10^6$ provided the size of the statistical ensemble is sufficiently large. As a result even if the population is known to be under the HW conditions there is an irreducible probability of approximately 0.15 that the $X^2$ estimator leads to wrong conclusions. Similar results are obtained for the case of non random mating. The ensemble distribution $P(X^2)$ is very similar to the case of random mating with a pronounced exponential tail and the probability of false positive is $P(X^2<3.8414)=0.848$. Therefore if the population is subject to an external bias (not detectable by other means) resulting in non random mating there is a probability of approximately 0.85 to get measurements leading to wrong estimations.

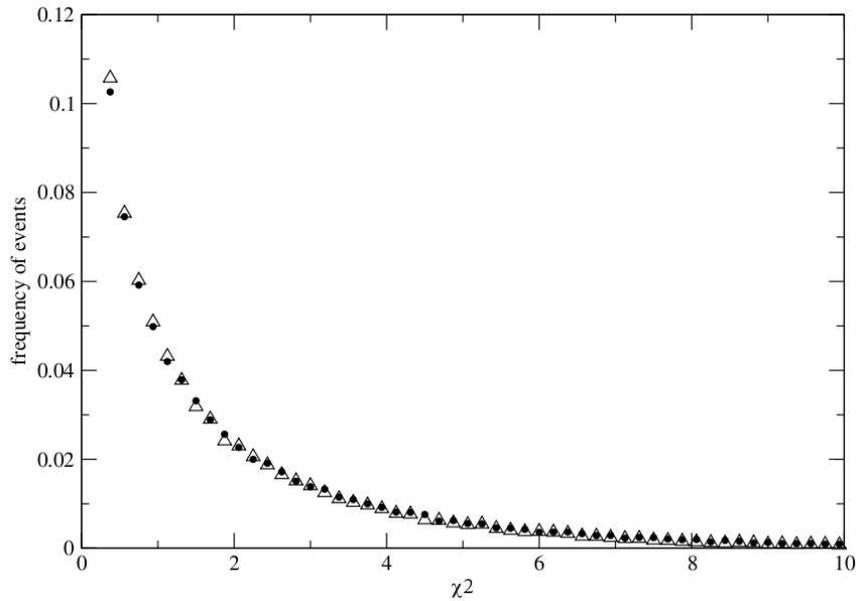

**Figure 2.** Distributions $P(X^2)$ over the statistical ensemble of $10^5$ populations composed by $3.0 \times 10^6$ individuals each. The values of $X^2$ are measured at fixed generation (simulation time unit) 30. The case of random mating (full dots) and the the case of nonrandom mating (empty triangles) are well fitted by exponential distributions for $X^2 > 0.4$ with a small deviation from the exponential function for . Both distributions are very stable with respect to the size of the statistical ensemble. The distributions are used to evaluate the probability of false negative (for the case of random mating) and the probability of false positive (for the case of nonrandom mating).

As a conclusion, the $X^2$ criterion can only be considered as a poor estimation of the conditions (HW or not) under which the genetic system is submitted. In other words, from the simple observation that $X^2 < 3.8414$ the existence of equilibrium states due to natural causes/mechanisms can not be logically sustained.

Another important point relates to the total number of reproducing couples that are typically formed in a given population at each reproduction period. The random mating condition is achieved as the larger the size of the population is and at the same time if the number of reproducing individuals is sufficiently small relative to the total population. In the numerical simulations this condition was satisfactorily fulfilled if the reproducing population was at most around 60% of the total population and if the mean reproductive capacity was sufficiently large in order to guarantee non extinction. In fact, from the strictly statistical point of view, if all the individuals reproduce then the most delayed formed couples (during the reproductive period) are necessarily biased due to sampling effects; the mating probabilities for the late couples become unbalanced due to small population size effects; individuals of the same genotype mate with different probabilities during the same reproduction period. The overall effect of having 100% mating can be simulated numerically observing the decrease of the probability of the allele fixation in finite time. In other words the most we approach the statistical conditions of random mating by fixing a much smaller percentage of the population as reproducing, the highest the probability to fix the allele in finite time due to random genetic drift. It is important to note that in spite of having total or partial mating in the population measured values of $X^2 < 3.8414$ are easily observed in both cases.

Finally it is important to highlight the fact that the $X^2$ test is based on the frequentist approach to probabilities that is unavoidably related to ideal (and frequently unattainable) situations. When applied to real situations this approach faces serious limitations because the analysis framework has to contemplate a set of frequently unverifiable assumptions about natural mechanisms that have to be assumed *a priori*. In the present case of genetic systems, as in many others, it is intuitive that the Bayesian approach would lead the analysis to more sound conclusions. Recent studies are in this perspective (Engels, 2009). By adopting this approach the "if then" strict relation between the HW axioms and the $X^2$ test has to be abandoned and therefore the notion of statistical equilibrium has to be considered in different terms. In this perspective the meaningful relation that contains the relevant question is

$$P(random\ mating | X^2 < 3.8414) = P(X^2 < 3.8414 | random\ mating) \cdot \frac{P(random\ mating)}{P(X^2 < 3.8414)}$$

**Discussion**

The strong limitations of the $X^2$ test to identify and characterize equilibrium states of the genetic system forces us to examine the concept of equilibrium in a deeply by appealing to its dynamical aspects. In fact, the relevant aspect of characterizing equilibrium states relates to the very dynamical properties of the system inscribed in its time evolution.

When the notion or concept of equilibrium is used as part of theoretical frameworks it becomes a crucial point to characterize the type(s) of equilibrium (equilibria) the natural system (or the model) has. In fact, the full characterization of the different equilibria as a function of the system's parameters may be seen as a reliable portrait of the model encoding the very essence of the principles governing the system. A coarse classification defines three types of equilibria: stable, neutral and unstable. The canonical way to identify the nature of the equilibrium state is to describe the system's behavior close to (in a small neighborhood of) the equilibrium state by means of small perturbations. For instance, in simple mechanical systems the case of a small rigid ball sliding on a concave surface under the action of the gravitational field and a small friction force is an example of a system that has a stable equilibrium state (the position where the ball has minimum potential energy). If the surface is convex the point of maximal potential energy is an unstable equilibrium state and in the case of a small ball on a flat and horizontal surface every point on the surface is a neutral equilibrium state of the mechanical system. Here the small ball constitutes the physical system and the surface is external to it and may be seen as the environment. The equilibrium state is an attribute of the system as a result of its interaction with its environment. In all three cases a fundamental property of equilibrium states is the time invariance under the action of the dynamic law; if $E$ is the identified equilibrium state and $S$ represents the dynamics then formally equilibrium states satisfy the relation $S(E)=E$.

For systems whose states are given by appropriate distribution functions of suitable observables the same concepts may be applied; one may talk about neutral, stable or unstable equilibrium states that are essentially statistical as a result of the intrinsic stochastic nature of the system. A simple example of stable (collective) states is described by distribution of velocities of molecules of a (non reacting) gas in thermal equilibrium with a heat bath; small perturbations on the gas may shift the system out of its equilibrium state but the system spontaneously goes back to its initial (stable equilibrium) state.

For genetic systems under the HW conditions the allele frequencies are fixed from the initial point $(t=0)$ and remain constant for the rest of the future system's history provided that the population is infinite. Small perturbations on the initial genotype frequencies will affect the time evolution by shifting the system to a new set of HW proportions which will remain constant in time with new values. This means that the genetic system under the HW conditions is better described in terms of neutral equilibrium states. In the appropriate jargon one should say "genetic systems under the HW conditions evolve according to neutral dynamics". This neutral aspect of the dynamics is the mechanism that allows genetic systems of finite size to evolve towards the fixation of alleles by random genetic drift.

If a genetic population is bounded in size (due to the balance between birth and death rates) and if the remaining HW conditions are respected, at each time step of observation (generation) the system is in a statistical state characterized by HW proportions. The change from one state to a new one is driven by the small fluctuations of the genotype frequencies which constitute a source of allele frequency changes internal to the system. Here the random mating condition is respected at each time step but mating probabilities for each genotype fluctuate along the system's history together with the genotype frequencies. Neutrality is a very special property of equilibrium states. In fact, it is the limiting case between stability and instability. Moreover, under HW conditions the genetic system has an infinite (uncountable) number of neutral equilibrium states labeled by all the values $0<f_a=1-f_A<1$ of the allele frequencies. This uncountable proliferation of neutral equilibria has a deep dynamical significance. It implies that the long term behavior of the system (measured in terms of the asymptotic time invariant values of the allele frequencies) may be as diverse as all the possible values $0<f_a<1$. As a consequence this asymptotic state depends on the genotype frequencies at $t=0$. In **Figure 3** we show the allele frequency time series of a numerical experiment where the population grows geometrically till the value $N=3.0 \times 10^7$ starting with the genotype profile $N_{AA}=10, N_{Aa}=20, N_{aa}=10$; random mating is considered at each time step.

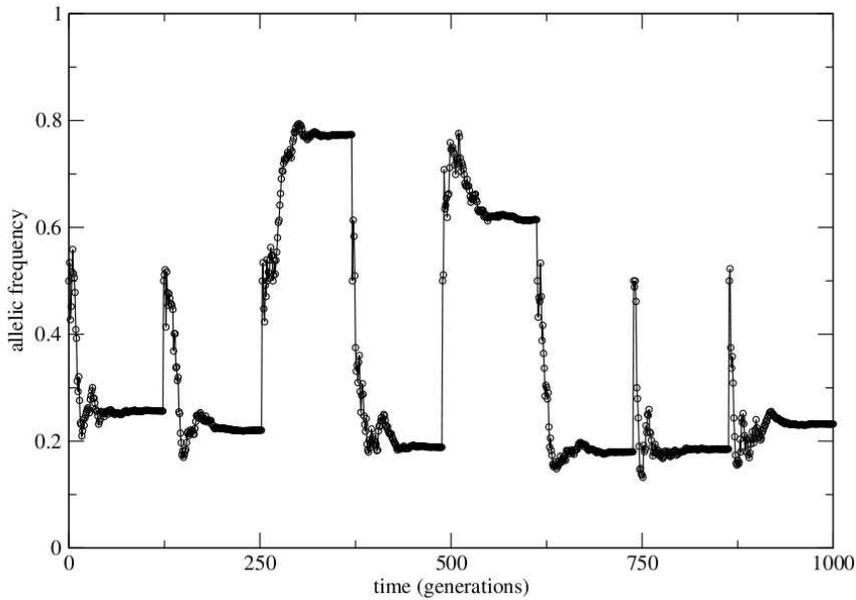

**Figure 3.** Time series of the allele-a frequency for the case of random mating. To highlight the role of neutral dynamics an initial population with the profile $N_{aa}=N_{AA}=10$ and $N_{Aa}=20$ (allele frequency $f_a=0.5$ ) grows till a fixed size of $N=3.0\text{x }10^7$ individuals. After reaching this maximal population size the allele frequency $f_a$ reaches approximately constant values interrupted by subsequent imposed bottlenecks reducing the population to its initial profile. After each bottleneck event a new transient takes place leading to new (approximately constant) value of $f_a$ due to the neutral characteristic of the dynamics.

The time necessary to reach the maximal population size is a transient period (around 100 generations) during which the population is subjected to fluctuations in the genotype frequencies. As the population grows these fluctuations decrease and the allele frequencies reach an approximate stationary value. Subsequent bottlenecks are imposed to the population bringing it to its initial configuration. As can be easily observed the approximately stationary values of the allele frequency are different after each bottleneck; the asymptotic behavior of the allele frequencies depends on the transient. This is a direct consequence of the neutral character of dynamics of finite populations under the HW conditions.

This scenario changes dramatically if the bounded population is conditioned by fixed *a priori* stable probabilities of mating relaxing the random mating HW condition. In this situation the system is driven by stable probabilities fixed by sources necessarily external to the system (note that from the logics of the model construction any information source out of the set of alleles $(A,a)$ is formally considered as external to the system). The overall effect over the time behavior of allele frequencies is time invariance with stable properties. Now small perturbations on the genotype frequencies rapidly die out in time and the initial (unperturbed) equilibrium state is restored. As it can be seen in **Figure 4** the initial small population (identical to the case of **Figure 3**) passes by a transient of approximately 100 generations and stabilizes its allele frequency around the asymptotic value $f_a=0.5$ . Just after stabilization of the allele frequency the population is subjected to a bottleneck reducing its size to its initial value; a new transient takes place and the allele frequency converges to the same previous asymptotic value $f_a = 0.5$. This convergence to the same value is the signature of stable dynamics.

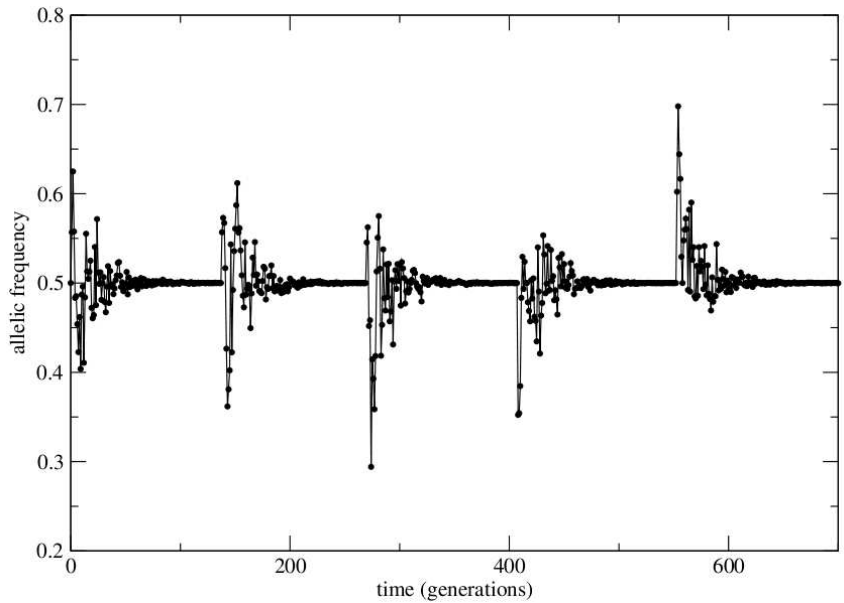

**Figure 4.** Time series of the allele-a frequency for the case of nonrandom mating. The initial and the after bottleneck sizes of the population and its maximal value are the same as in the case of Figure 3. After each bottleneck event the population experiences a transient period of approximately 100 generations and then the allele frequency stabilizes at $f_a = 0.5$. The subsequent bottleneck events imposed to the population shows the robustness of the transient size and of the asymptotic value of the allele frequency due to stable character of the dynamics.

It is important to note that typical values of $X^2$ are remarkably small if the equilibrium state of the system is either neutral or stable. Clearly under unbalanced mating probabilities the system becomes trapped on a stable state and fixation of alleles is impossible. The fact that the genetic system may evolve under neutral or stable dynamics (depending on the balanced or unbalanced nature of the mating probabilities) suggests the following plausible scenario: in real situations the system is always finite and as a result of many interdependent variables the system could be subject to a time interplay of balanced and unbalanced mating probabilities. In this case the allele frequency time series would be typically a sequence of time periods of random drift separated by periods of stable allele frequencies in such a way that observed values of $X^2$ are typically small at any time. Depending on how frequently the system jumps between the balanced and the unbalanced context (which would depend on additional dynamical rules) the properties of equilibrium states could be hardly identified. One may easily conceive a dynamical portrait where different stable and neutral equilibrium states dynamically appear and disappear leading to a concept close to the idea of metastable states that are locally (in a small interval of the allele frequencies) but not globally (for all values of the allele frequencies) stable. Therefore, the structure of the dynamics in genetic systems may be very subtle, complex and hardly identifiable using simple tools like the $X^2$ test. As a conclusion the notion of equilibrium states, absent in G.

Hardy original letter (Hardy, 1908), deserves deeper analysis well beyond the simplistic discussions found in textbooks and papers.

In several cases the problem of characterizing equilibrium states is very complex to be treated analytically. In these cases it is usual to have only the output time series of relevant observables as workable data and the important quantity to be examined is the time auto correlation function of the relevant observable. For systems driven by stable equilibrium states this function presents fast (exponential like) decay to zero with a characteristic decay rate that relates to the typical time necessary to achieve the equilibrium (asymptotic) state; the time to achieve equilibrium is an intrinsic property of the system and is usually called the system's relaxation time. In this perspective exponential decay of correlation functions is viewed as a signature of stable dynamics and an indirect way to identify the existence of stable equilibrium states. In other terms, stable dynamics leads to (fast enough) exponential decay of correlations characterizing short transients and stable equilibrium states achievable in finite time. That is the case of genetic systems subject to constant (external) unbalanced mating probabilities: the auto correlation function for the allelic frequency (considered as the system's observable) typically decays to zero in a exponential way: this is the process of relaxation towards equilibrium. On the other hand, in the case of random mating the decay of the allelic frequency auto correlation is typically (slow) non exponential characterizing a system driven by neutral dynamics. The decay in this cases is so slow that the system never attains an equilibrium state. The system is not driven by an equilibrium state but its dynamics is governed by an endless transient regime.

One main implication of this fact is that if the system's auto correlation function does not relax fast enough then the (eventual) existence of an equilibrium state in the limit of infinite time becomes useless in helping to understand and describe the system's dynamics. The whole dynamics is contained in the transient regime and the time history of the systems is better described as a permanent transformation without asymptotic equilibrium. That is the case of systems evolving according to neutral dynamics. If the system is rigorously infinite then the observation of transient times is impossible since the system is in neutral equilibrium from the starting point $t=0$.

On the other hand if the system is finite (even very large) the transient regime is infinite and the system never attains a state that possibly could be classified as equilibrium unless we define it as such theoretically with no way to make it unambiguously observable. Two representative examples are given in **Figures 5 and 6**. In **Figure 5** the system evolves according to random mating and the slow decay of correlations is visible if compared with **Figure 6** showing the same allele frequency time correlation function for one case of non random mating.

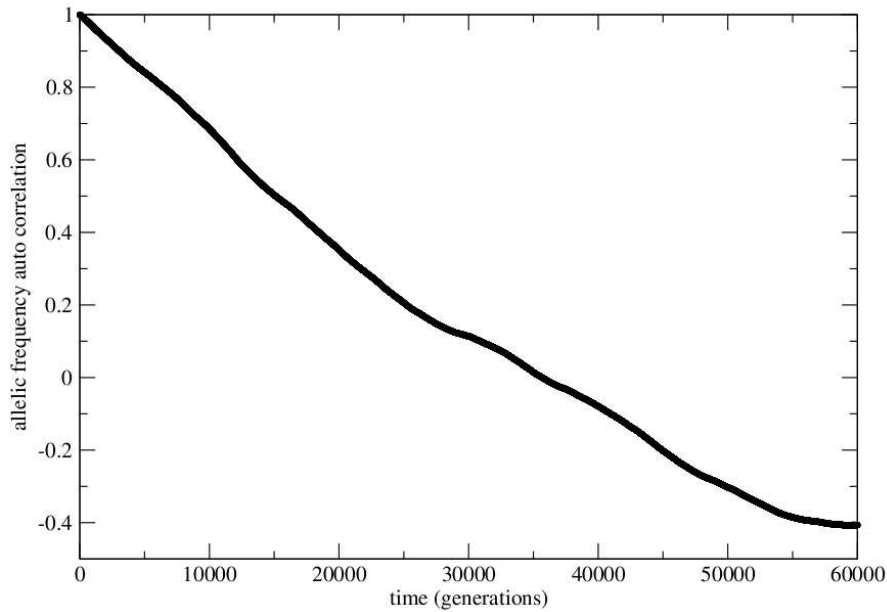

**Figure 5.** The allelic frequency auto correlation function for the case of random mating. The data were obtained from the allele frequency time series for a population of $3.0 \times 10^6$ individuals during a history of $3.0 \times 10^5$ generations. The auto correlation function shows that the genetic system does not reach an equilibrium state till at least 60,000 generations.

In **Figure 6** the decay is typically exponential and the correlation function can be written as $C(t) \approx \exp(-dt)$ with $T = 1/d$ the characteristic time needed to achieve the equilibrium state; in other words the correlation function decays fast enough such that the equilibrium state is observable in finite time. On the other hand in **Figure 5** the decay is slower than any exponential function in such a way that the time needed to achieve equilibrium is beyond observability (in the case of the present model rigorously infinite). This observation leads to the conclusion that under the HW conditions the genetic population is permanently relaxing towards an unattainable state of equilibrium. It should be clear that after fixing one allele the population is in a state easily identifiable as stable equilibrium since small perturbations of this state would easily die out restoring the state of fixed allele.

The relevance of the above discussion is based on addressing the fundamental dynamical aspect of the concept of equilibrium (states) in relation to the idea that under HW conditions genetic populations show time invariance of the allele frequencies. Therefore, to characterize the equilibrium states is a central issue. In order to identify the properties of the equilibrium state revealed by the system's time series one should apply dynamical criteria and not statistical ones. Experimental studies attempting to identify the properties of equilibrium states should try to reproduce what is shown is **Figures 3, 4, 5 and 6** namely: if the same or different allele proportions are obtained after introducing perturbations and/or how allele time correlations decay for systems with partial mating. This would offer sound arguments for the hypothesis of equilibrium and its properties. Moreover, we note that to fully justify the hypothesis of equilibrium states one should take in full consideration the biological/physical mechanisms involved in reproduction and their impact on the system's history.

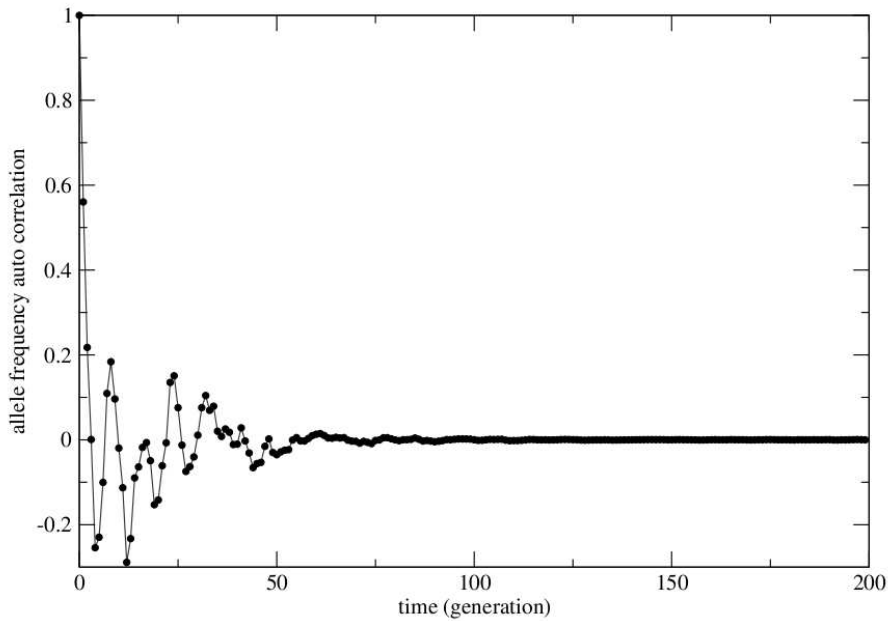

**Figure 6.** The allelic frequency auto correlation function for one representative case of nonrandom mating. In order to show the transient regime plus the achievement of equilibrium the numerical simulation was performed with an initial population $(N_{aa}=10, N_{AA}=10, N_{Aa}=20)$ that grows till a population size of $3.0\text{x}10^7$ individuals. The simulation total time is $3.0\text{x}10^5$ generations. After approximately 100 generations the correlation function is very close to zero indicating that (stable) equilibrium has been reached.

Analogies between theories should be grounded on clear and sound basis going as far as possible to the foundations, otherwise it is very likely that misleading conclusions will be derived from such analogies. The attempt to apply the notion of equilibrium to genetic systems from the analogous concept in mechanics is one of these cases. The notion of a genetic state of equilibrium presented in the context of the synthetic theory implies necessarily that the driving cause of genetic transformation of the system's state is external to it in analogy to the action of external forces acting on a mechanical system through one or more of the basic physical interactions. Therefore, the analogy here is established at the level of Newton's second law which is a definition of force. The analogy would only be complete if one is capable to derive or identify a principle analogous to the Newton's third law which contains the physical essence of the principles governing mechanics. If we conceive such a principle applied to genetic systems (which finally would describe how genetic systems interact with their environment) we are obliged to consider the influences due to the size of the environment in respect to the system's size as well as the specific role of what could be external and internal forces and their nature as conservative or dissipative. The distinction between these two forces is deeply related to the existence of conservation laws and time symmetry. The general difficulty to identify a version of the Action-Reaction principle for genetic systems and the strong indications that stochasticity plays a fundamental role in genetic dynamics (not only due to sampling limitations) imposes the identification of genetic driving forces as essentially dissipative. This point is supported by the fact that genetic populations are systems with varying total mass and therefore have to be considered as open physical systems.

The above arguments together with the fact that genetic systems are usually composed by a large number of individuals suggests that a better formal analogy should be made with non-equilibrium statistical physics rather than with mechanics which would focus on the search of basic principles related to the second law of thermodynamics or the flux of fundamental quantities like entropy. In dealing with open systems we have the conceptual advantage that independently of the chosen driving (extreme) principle it is quite natural to expect the theory could describe the existence of stationary states that may be classified as stable (locally or globally), metastable, unstable or even situations where a large number of different stationary states may coexist leading to very complex dynamical portraits. As it should be clear at this point, in pushing the analogy with non-equilibrium thermodynamics and statistical physics we are forced to abandon the analogy with mechanical equilibrium and consider the idea of genetic equilibrium states as stated by the HW principle just as a unattainable idealization in the same way closed isolated thermodynamic systems are idealized objects with no counterpart in reality. But here again the analogy is not complete: ideal thermodynamic systems are of great importance to put in evidence the existence of general laws governing their dynamical behavior (the laws of equilibrium thermodynamics and the emergence of dissipative structures out of equilibrium). For genetic systems this is not the case so far.

**Acknowledgements**

The authors thank Dr. Maximo Sandin and Dr. Mauricio Abdalla for careful reading and critical comments on the manuscript. This work was supported by Grants from FAPESP, Brazil to F.B. And by FAPESP, CNPq and CAPES, Brazil to M.R.S.B.